\documentclass{jps-cp}
\usepackage{txfonts} %Please comment out this line unless the txfonts package is availabe in your LaTeX system.
\usepackage{color}

\def\gsim{\mathop {\vtop {\ialign {##\crcr 
$\hfil \displaystyle {>}\hfil $\crcr \noalign {\kern1pt \nointerlineskip } 
$\,\sim$ \crcr \noalign {\kern1pt}}}}\limits}
\def\lsim{\mathop {\vtop {\ialign {##\crcr 
$\hfil \displaystyle {<}\hfil $\crcr \noalign {\kern1pt \nointerlineskip } 
$\,\,\sim$ \crcr \noalign {\kern1pt}}}}\limits}

\title{Quantum Valence Criticality in Heavy Fermions 
on Periodic and Aperiodic Crystals}

\author{Shinji \textsc{Watanabe}$^{1}$ and Kazumasa \textsc{Miyake}$^{2}$}

\inst{$^{1}$Department of Basic Sciences, Kyushu Institute of Technology, Kitakyushu, Fukuoka 804-8550, Japan \\
$^{2}$Center for Advanced High Magnetic Field Science, Osaka University, Toyonaka, Osaka 560-0043, Japan}

\email{swata@mns.kyutech.ac.jp}

\recdate{September 8, 2019}

\abst{Progress of theories and experiments of the quantum valence criticality is overviewed focusing on recent discoveries of direct evidence of quantum critical point of valence transitions in periodic crystal $\alpha$-YbAl$_{1-x}$Fe$_x$B$_4$ at $x=0.014$ and quasicrystal Yb$_{15}$(Al$_{1-x}
$Ga$_{x}$)$_{34}$(Au$_{1-y}$Cu$_{y}$)$_{51}$ at $x=y=0$.  
The common quantum criticality as well as $T/B$ scaling behavior in the magnetization observed in the periodic crystal $\beta$-YbAlB$_4$ and quasicrystal Yb$_{15}$Al$_{34}$Au$_{51}$ is shown to be explained from the theory of critical Yb-valence fluctuations starting from the Hamiltonians for describing the low-energy electronic states in both systems. 
}

\kword{Quantum valence criticality, quantum critical point of valence transition, $T/B$ scaling, $\beta$-YbAlB$_4$, quasicrystal Yb$_{15}$Al$_{34}$Au$_{51}$,   $\alpha$-YbAl$_{1-x}$Fe$_x$B$_4$, approximant crystal Yb$_{14}$Al$_{35}$Au$_{51}$}

\begin{document}
\maketitle

\section{Introduction}

Quantum critical phenomena have attracted much attention in condensed matter physics. 
When continuous-transition temperature to ordered phase is suppressed to absolute zero by changing control parameters of materials such as pressure, magnetic field, and chemical substitution, a quantum critical point (QCP) is realized. 
Near the QCP, physical quantities exhibit non-Fermi-liquid behaviors at low temperatures, which are called quantum critical phenomena. 
Quantum critical phenomena emerging near the magnetic QCP have been well understood from the theory of 
spin fluctuations~\cite{Moriya} and the renormalization-group theory~\cite{Hertz,Millis}. 
The quantum criticality near the antiferromagnetic (AF) QCP in three spatial dimension (3D) 
is shown in the first line of Table~\ref{tb:QCP}.  

However, unconventional quantum criticality has been observed in several heavy-fermion materials such as 
YbCu$_{3.5}$Al$_{1.5}$~\cite{Bauer}, YbRh$_2$Si$_2$~\cite{Trovarelli,Gegenwart}, $\beta$-YbAlB$_4$~\cite{Nakatsuji,Matsumoto}, as shown from the second to fourth lines of Table~\ref{tb:QCP}, which has been one of the central issues in strongly correlated electron systems.  
As a possible origin, the theory of critical valence fluctuations 
arising from the QCP of the valence transition has been proposed, which gives a unified explanation for the unconventional quantum critical phenomena (see the last line of Table~\ref{tb:QCP})~\cite{WM2010}. 

Interestingly, a new heavy-fermion metal Yb$_{15}$Al$_{34}$Au$_{51}$ with the quasicrystal (QC) lattice structure has been synthesized, which exhibits the common quantum criticality to that in $\beta$-YbAlB$_4$, as shown in the second last line of Table~\ref{tb:QCP}~\cite{Deguchi,Watanuki}.
Recently, direct evidence of the valence QCP has been observed experimentally in the periodic crystal $\alpha$-YbAl$_{1-x}$Fe$_x$B$_4$ at $x=0.014$~\cite{Kuga2018} and QC Yb$_{15}$(Al$_{1-x}
$Ga$_{x}$)$_{34}$(Au$_{1-y}$Cu$_{y}$)$_{51}$ at $x=y=0$~\cite{Imura}. 
These observations evidence that the QCP of the valence transition is realized in the periodic crystal $\beta$-YbAlB$_4$ and QC Yb$_{15}$Al$_{34}$Au$_{51}$. 

In this paper, we overview the recent progress of theories and experiments of quantum valence criticality in heavy fermions on periodic and aperiodic crystals focusing on these measurements. 
We discuss that the common quantum criticality i.e., the exponent $\zeta=1/2$ in Table~\ref{tb:QCP} as well as a new type of scaling called $T/B$ scaling is theoretically shown to appear in the periodic crystal $\beta$-YbAlB$_4$ and QC Yb$_{15}$Al$_{34}$Au$_{51}$ starting from the Hamiltonians for describing the low-energy electronic states in both systems. 

The organization of this paper is as follows. 
Quantum critical phenomena in $\beta$-YbAlB$_4$ and QC Yb$_{15}$Al$_{34}$Au$_{51}$ are discussed 
in Sect.~\ref{sec:b_YbAlB4} and Sect.~\ref{sec:QC}, respectively. 
Summary and perspective are given in Sect.~\ref{sec:SP}.

%%%%%%%%%%%%%%%%%%%% Table 1 %%%%%%%%%%%%%%%%%%%%%%%%%%%%%%%%%%%%%%%%%%%%%%%%%%%%%%%%
\begin{table}[tb]
\caption{Theories and materials and their low-temperature properties of uniform magnetic susceptibility $\chi$, 
 specific-heat coefficient $C/T$, 
resistivity $\rho$, and NMR/NQR relaxation rate $(T_{1}T)^{-1}$. 
As for valence criticality, 
$T> T_0$ regime with $T_0$ being 
characteristic temperature of critical valence fluctuations is shown (as for resistivity, see Sect.~\ref {sec:res} for detail). 
$\zeta$ takes the value for 
$0.5\le\zeta\le0.7$ depending on materials and temperature range (see \cite{WM2010} for detail). 
In $\beta$-YbAlB$_4$ and Yb$_{15}$Al$_{34}$Au$_{51}$, $\zeta=1/2$ is shown (see Sects~\ref {eq:TB} and \ref{sec:AC_TB}, respectively). }
\label{tb:QC}
\begin{center}
\begin{tabular}{lccccc}
\hline
\multicolumn{1}{c}{Material/Theory} & \multicolumn{1}{c}{$\chi(T)$}  & 
\multicolumn{1}{c}{$C(T)/T$} & \multicolumn{1}{c}{$\rho(T)$} & \multicolumn{1}{c}{$(T_{1}T)^{-1}$} & Refs. \\
\hline
%Fermi liquid & const. & const. & $T^2$ & const. &  \\
3D AF criticality & const.$-T^{1/4}$ & const.$-T^{1/2}$ & $T^{3/2}$ & $T^{-3/4}$ & \cite{MT,HNM} \\
YbCu$_{3.5}$Al$_{1.5}$ & $T^{-2/3}$ & $-\ln{T}$ & $T^{1.5} \ \to \ T$ & * & \cite{Bauer} \\
YbRh$_2$Si$_2$ & $T^{-0.6}$ & $-\ln{T}$ & $T$ & $T^{-0.5}$ & \cite{Trovarelli,Gegenwart} \\
$\beta$-YbAlB$_4$ & $T^{-0.5}$ & $-\ln{T}$ & $T^{1.5} \ \to \ T$ & * & \cite{Nakatsuji,Matsumoto} \\
$\alpha$-YbAl$_{0.086}$Fe$_{0.014}$B$_4$ & $T^{-0.5}$ & $-\ln{T}$ & $T^{1.6} \ \to \ T$ & 
%%%%%
%\textcolor{red}
%{
$T^{-0.5\sim -0.7}$
%}
%%%%% 
& \cite{Kuga2018,
MacLaughlin
} \\
Yb$_{15}$Al$_{34}$Au$_{51}$ & $T^{-0.51}$ & $-\ln{T}$ & $T$ & $\propto\chi$ & \cite{Deguchi} \\
Quantum valence criticality & $T^{-\zeta}$ & $-\ln{T}$ & $T$ &  $\propto\chi$ & \cite{WM2010} \\
\hline
\end{tabular}
\label{tb:QCP}
\end{center}
\end{table}
%%%%%%%%%%%%%%%%%%%%%%%%%%%%%%%%%%%%%%%%%%%%%%%%%%%%%%%%%%%%%%%%%%%%%%%%%%%%%%%%%%%%%

\section{Quantum criticality in periodic crystal $\beta$-YbAlB$_4$}
\label{sec:b_YbAlB4}

\subsection{Direct evidence of valence QCP in $\alpha$-YbAl$_{1-x}$Fe$_x$B$_4$~$(x=0.014)$}
\label{sec:a_YbAlFeB4}

The heavy-fermion metallic compound 
$\alpha$-YbAlB$_4$, which is a sister compound of $\beta$-YbAlB$_4$, shows the Fermi-liquid behavior. However, by replacing Al with Fe $1.4\%$, the same quantum criticality as that in $\beta$-YbAlB$_4$ appears~\cite{Kuga2018}. 
Furthermore, at $x=0.014$, the $T/B$ scaling behavior where magnetic susceptibility is expressed as a single scaling function of the ratio of temperature $T$ and magnetic field $B$ appears~\cite{Kuga2018}. 
The $T/B$ scaling was also observed in $\beta$-YbAlB$_4$~\cite{Matsumoto}. 
Interestingly, at just $1.4\%$, the sharp change in Yb valence as well as the volume change has been observed, as shown in Fig.~\ref{fig:a_YbAlFeB4}(a). This is the direct observation of the sharp valence crossover line arising from the valence QCP in Fig.~\ref{fig:a_YbAlFeB4}(b) with quantum valence criticality.
This observation verifies the theory of critical valence fluctuations in Ref.~\cite{WM2010}.

\subsection{$T/B$ scaling and $\chi(T)\sim T^{-0.5}$ in $\beta$-YbAlB$_4$}
\label{eq:TB}

The $T/B$ scaling is also explained by the theory of critical valence fluctuations based on the Hamiltonian for describing the low-energy electronic state in $\beta$-YbAlB$_4$~\cite{WM2014}. 
Since the crystalline-electric-field (CEF) ground state of 4f hole at Yb in $\beta$-YbAlB$_4$ is $|J=7/2,\pm 5/2\rangle$, 
%~\cite{NC2009}, the hybridization between the Yb 4f electron and the B 2p electron is considered to be anisotropic~\cite{RC2014}. 
%%%%%
%\textcolor{red}
%{
the hybridization between the Yb 4f electron and the B 2p electron is shown to be anisotropic, which has a line node along the $c$ direction~\cite{NC2009}. 
%}
%%%%%

The effective Hamiltonian in the hole picture is given by the extended periodic Anderson model 
under the magnetic field: 
\begin{eqnarray}
H=\sum_{{\bf k}\sigma}\varepsilon_{\bf k}
c_{{\bf k}\sigma}^{\dagger}c_{{\bf k}\sigma}
+\varepsilon_{\rm f}\sum_{i\sigma}n_{i\sigma}^{\rm f}
+\sum_{{\bf k}\sigma}\left(V_{\bf k}
f_{{\bf k}\sigma}^{\dagger}c_{{\bf k}\sigma}
+{\rm h.c.}
\right)
+U\sum_{i}n_{i\uparrow}^{\rm f}n_{i\downarrow}^{\rm f}
+U_{\rm fc}\sum_{i\sigma\sigma'}n_{i\sigma}^{\rm f}n_{i\sigma'}^{\rm c}
-h\sum_{i}
S_{i}^{{\rm f}z}, 
\label{eq:H_bYbAlB4}
\end{eqnarray}
where the conduction holes with the conduction band $\varepsilon_{\bf k}$ hybridizes with the 4f hole with the 4f level $\varepsilon_{\rm f}$ via the hybridization $V_{\bf k}$. 
Here, $\sigma=\uparrow, \downarrow$ denote the Kramers pairs. 
To simulate the anisotropic hybridization in $\beta$-YbAlB$_4$, we take $V_{\bf k}=V(1-\hat{k}_{z}^{2})$ 
with $\hat{\bf k}\equiv{\bf k}/|{\bf k}|$~\cite{NC2009,IM1996}. 
The 4th term is the onsite Coulomb repulsion between 4f holes $U$  
and the 5th term is the Coulomb repulsion between 4f and conduction holes $U_{\rm fc}$ 
with $n_{i\sigma}^{a}\equiv a_{i\sigma}^{\dagger}a_{i\sigma}$ 
for ${\rm a}={\rm f}$ or ${\rm c}$. 
%%%%%
%\textcolor{red}
%{
Hereafter, the unit of energy is taken as the half bandwidth of the conduction band unless otherwise stated.
%}
%%%%%
The last term is the Zeeman term for 4f holes where $h$ is the magnetic field and $S_{i}^{{\rm f}z}$ is defined by $S_{i}^{{\rm f}z}\equiv \frac{1}{2}(n_{i\uparrow}^{\rm f}-n_{i\downarrow}^{\rm f})$. 

Starting from the Hamiltonian Eq.~(\ref{eq:H_bYbAlB4}), we derived the mode-mode coupling equation of the critical valence fluctuations, i.e., self-consistent renormalization (SCR) equation for valence fluctuations (see Ref.~\cite{WM2014} for details).  
By solving the valence SCR equation at the valence QCP, 
we obtain the solution $y$. 
Here, $y$ is proportional to the inverse valence susceptibility $y\propto\chi_{\rm v}^{-1}$~\cite{WM2014}.
Since the uniform magnetic susceptibility $\chi$ also diverges at the valence QCP, which is caused by the common many-body effect to the critical valence fluctuations caused by $U_{\rm fc}$ in Eq.~(\ref{eq:H_bYbAlB4}), the uniform magnetic susceptibility is proportional to the valence susceptibility near the valence QCP so that $y\propto\chi_{\rm v}^{-1}\propto\chi^{-1}$ holds~\cite{WM2008,WM2010}.

\begin{figure}[tb]
\includegraphics[width=15cm]{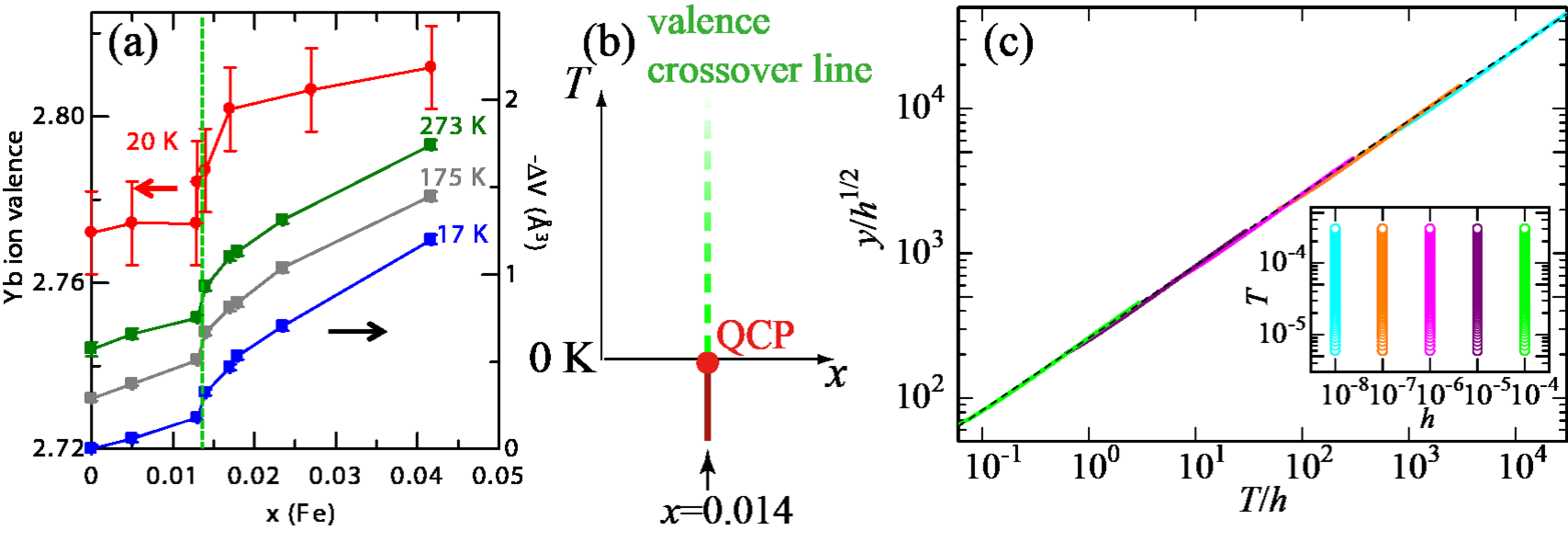}
\caption{(Color online) (a) Fe-concentration dependence of Yb valence (left axis) and volume change (right axis) in $\alpha$-YbAl$_{1-x}$Fe$_x$B$_4$~\cite{Kuga2018}. 
(b) Phase diagram of $\alpha$-YbAl$_{1-x}$Fe$_x$B$_4$ in the plane of temperature $T$ and Fe concentration $x$. First-order valence transition (solid line) terminates at the QCP (filled circle), from which  a sharp valence crossover line extends (dashed line). 
(c) Scaling plot of $y/h^{1/2}$ vs. $T/h$. Here, $y(T, h)$ is the solution of the valence SCR equation where the parameters except for $T$ and $h$ are set for the valence QCP at $h=0$~\cite{WM2014}. 
The dashed line is the least-square fit of the data (see text). 
The inset shows the data in the $T$-$h$ plane where the scaling applies. 
} 
\label{fig:a_YbAlFeB4}
\end{figure}

The result is shown in Fig.~\ref{fig:a_YbAlFeB4}(c). 
We find that all the data for the temperature range and magnetic-field range shown in the inset corresponding to the experimental-data range of 
%$T<3$~K and $B<2$~T 
%%%%%
%\textcolor{red}
%{
$3\times 10^{-2}$~K$\le T\le 3$~K and $10^{-4}$~T$\le B\le 2$~T 
%}
%%%%%
fall down to a single scaling function of the ratio of $T$ and $h$  
\begin{eqnarray}
y=h^{1/2}\phi\left(\frac{T}{h}\right), 
\label{eq:scaling}
\end{eqnarray}
where $\phi(x)=ax^{1/2}$ is obtained by the least-square fit for the large $T/h$ range indicated by the dashed line in Fig.~\ref{fig:a_YbAlFeB4}(c). 
Here, the unit of energy is taken as the half bandwidth of the conduction band and we set $k_{\rm B}$=1.   
Hence, $y/h^{1/2}=a(T/h)^{1/2}$ holds, which gives $y\propto T^{1/2}$. 
This result $\chi h^{1/2}=a'(h/T)^{1/2}$ is in accordance with the measured $T/B$ scaling behavior in $\beta$-YbAlB$_4$~\cite{Matsumoto} and also in $\alpha$-YbAl$_{0.086}$Fe$_{0.014}$B$_4$~\cite{Kuga2018}. 
This implies that the quantum valence criticality is dominant in this regime, 
giving rise to the non-Fermi-liquid regime. 
The above result indicates that the magnetic susceptibility as well as the valence susceptibility 
behaves as  
\begin{eqnarray}
\chi\propto T^{-1/2}, 
\label{eq:chi_bYbAlB4}
\end{eqnarray} 
for the zero-field limit, $h\to 0$, which explains the measured criticality 
in $\beta$-YbAlB$_4$ (see Table~\ref{tb:QCP}). 

%%%%%%%%%%%%%%%  Fig.1 %%%%%%%%%%%%%%%%%%%%%%%%%%%%
%%%%%%%%%%%%%%%%%%%%%%%%%%%%%%%%%%%%%%%%%%%%%%%%%%%

\subsubsection{Very slow fluctuations of Yb valence in $\beta$-YbAlB$_4$}
\label{sec:small_T0}

By analyzing the valence SCR equation under magnetic field, it turned out that the $T/B$ scaling appears in case that the characteristic temperature of critical valence fluctuations $T_0$ is comparable to or less than the measured lowest temperature~\cite{WM2014}. 
Indeed, the result in Fig.~\ref{fig:a_YbAlFeB4}(c) was obtained for $T_0=3\times 10^{-6}$, which is located around the lowest temperature [see the inset in Fig.~\ref{fig:a_YbAlFeB4}(c)]. 
The emergence of the extremely small $T_0$ was explicitly shown in the original paper~\cite{WM2010} on the basis of the extended periodic Anderson model. 
Because of the strong local correlation effect by large Coulomb repulsion $U$, almost flat mode of critical valence fluctuations appears around the $\Gamma$ point in the momentum space, giving rise to extremely small $T_0$. 
Then, a new type of quantum criticality shown in Table~\ref{tb:QCP} as well as the $T/B$ scaling in Fig.~\ref{fig:a_YbAlFeB4}(c) emerges. 

Recently, by M{\"o}ssbauer measurement in $\beta$-YbAlB$_4$, very slow valence-fluctuation time scale 
$\tau\approx 2$~ns has been observed~\cite{Kobayashi}. 
This corresponds to $T_0\approx 24$~mK, which is around the measured lowest 
temperature $T=30$~mK.
As noted in Sect.~\ref{sec:a_YbAlFeB4}, in $\alpha$-YbAl$_{1-x}$Fe$_x$B$_4$, the same quantum criticality and $T/B$ scaling as those in $\beta$-YbAlB$_4$ were observed at $x=0.014$. 
Hence, the very slow valence fluctuation similarly to $\beta$-YbAlB$_4$ is expected to be observed in $\alpha$-YbAl$_{0.086}$Fe$_{0.014}$B$_4$. 
Such a measurement is left for interesting future subject. 

\subsubsection{Resistivity in $\beta$-YbAlB$_4$ and $\alpha$-YbAl$_{0.086}$Fe$_{0.014}$B$_4$}
\label{sec:res}

 In $\beta$-YbAlB$_4$, 
the $T$-linear resistivity $\rho(T)\sim T$ is observed for $T\gsim 1$~K, while 
$\rho(T)\sim T^{1.5}$ is observed for $T\lsim 1$~K [in the linear plot of $\rho(T)$ the rounding seems to appear for $T\lsim 0.3$~K]~\cite{Nakatsuji}. 
Theoretically, 
it has been shown that the locality of the critical valence-fluctuation mode gives rise to the $T$-linear resistivity~\cite{HJM,WM2010}. 
Here we note that the quantum valence criticality, which differs from conventional magnetic criticality, appears
% for $T>T_{0}$
in the higher-$T$ regime than a few times of $T_0$~\cite{WM2010}. 
For $T<T_0$, resistivity behaves as $\rho(T)\sim T^{1.5}$ in the dirty system and $\rho(T)\sim T^{5/3}$ in the clean system. 
Hence, if we adopt $T_0=24$~mK in $\beta$-YbAlB$_4$, the rounding of the low-$T$ part of the resistivity is theoretically expected to appear below $T\lsim 5T_0\approx 120$~mK, as explicitly shown in Ref.~\cite{WM2010}. 
This is 
%actually the case observed 
favorably compared with the measured behavior of $\rho(T)$ 
in $\beta$-YbAlB$_4$~\cite{Nakatsuji}. 
Therefore, both behaviors $\rho(T)\sim T^{1.5}\to T$ listed for $\beta$-YbAlB$_4$ in Table~\ref{tb:QCP} are naturally explained. 
Furthermore, as for $\alpha$-YbAl$_{0.086}$Fe$_{0.014}$B$_4$, both behaviors $\rho(T)\sim T^{1.6}\to T$ in Table~\ref{tb:QCP} are also explained from the theory of critical valence fluctuations.  

%%%%%
%\textcolor{red}
%{
\subsubsection{Dynamical relaxation rate in $\alpha$-YbAl$_{0.086}$Fe$_{0.014}$B$_4$}
%}
%%%%%

%%%%%
%\textcolor{red}
%{
In $\alpha$-YbAl$_{0.086}$Fe$_{0.014}$B$_4$, the $\mu$SR measurement has recently detected the temperature dependence of the dynamical relaxation rate $\lambda_{\rm d}(T)\sim T^{0.3\sim 0.5}$, which starts to appear below 0.1~K~\cite{MacLaughlin}. Since $\lambda_{\rm d}$ is proportional to ${T_1}^{-1}$, this behavior is consistent with the quantum valence criticality, as shown in Table~\ref{tb:QC}. However, it was  necessary to explain why such a behavior starts to appear below $T\sim 0.1$~K. The explanation was given as follows~\cite{MW2018}: $\mu^{+}$ attracts conduction electrons, around it inducing the local magnetic moment on Yb ion causing the impurity Kondo effect, which in turn affects the relaxation of $\mu^{+}$.  Then, on cooling, the dynamical relaxation rate, $\lambda_{\rm d}\propto{T_1}^{-1}$ increases and shows the Koringa relation below the muon-induced local Kondo temperature $T_{\rm K}^{*}$, which is estimated to be around $0.2$~K. Then, in the temperature regime $T<T_{\rm K}^{*}$, where $\mu^{+}$ is subject to the CVF at the valence QCP directly, the anomalous $T$ dependence in ($T_{1}T)^{-1}\propto T^{-0.5~\sim -0.7}$ is manifested (see Table~\ref{tb:QC}). 
%}
%%%%%

%------------------------------------------------------------------------------------------
\section{Quantum criticality in quasicrystal Yb$_{15}$Al$_{34}$Au$_{51}$}
\label{sec:QC}

The QC Yb$_{15}$Al$_{34}$Au$_{51}$ exhibits the common quantum criticality to that in $\beta$-YbAlB$_4$, 
as shown in Table~\ref{tb:QCP}, which surprisingly persists even under pressure~\cite{Deguchi}.  Furthermore, essentially the same $T/B$ scaling behavior has been observed in the QC Yb$_{15}$Al$_{34}$Au$_{51}$~\cite{DS_pc,Matsukawa2014}. 
Recently, systematic variations of homologous elements of Au and Al in the QC Yb$_{15}$Al$_{34}$Au$_{51}$, i.e., Yb$_{15}$(Al$_{1-x}$Ga$_{x}$)$_{34}$(Au$_{1-y}$Cu$_{y}$)$_{51}$, have been synthesized by Imura et al~\cite{Imura}. Interestingly, in the plot of the Yb valence vs. the lattice constant, the QC Yb$_{15}$Al$_{34}$Au$_{51}$, which exhibits quantum criticality, is located at the point where the Yb valence starts to change sharply. This indicates that Yb valence plays a key role for the quantum criticality.
In Sect.~\ref{sec:QC_AC}, the geometry of the QC and the relation to the approximant crystal (AC) is briefly explained and the lattice-constant dependence of the Yb valence in the QC Yb$_{15}$(Al$_{1-x}$Ga$_{x}$)$_{34}$(Au$_{1-y}$Cu$_{y}$)$_{51}$ is analyzed theoretically in Sect.~\ref{sec:Lattice_QC}. 
Theoretical explanations for the quantum criticality as well as $T/B$ scaling and robustness of the quantum criticality are given in Sect.~\ref{sec:AC_TB} and Sect.~\ref{sec:robustP}, respectively. 

\subsection{Quasicrystal and approximant crystal}
\label{sec:QC_AC}

\begin{figure}[b]
\includegraphics[width=14cm]{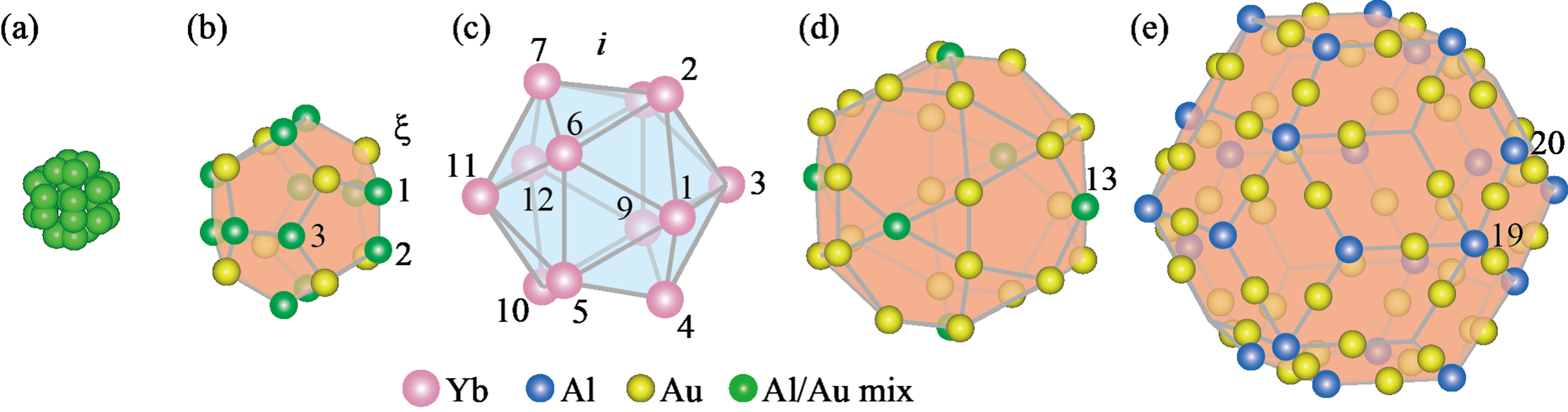}
\caption{(Color online) Concentric shell structures of the Tsai-type cluster in the approximant crystal Yb$_{14}$Al$_{35}$Au$_{51}$~\cite{Deguchi,Ishimasa}:
(a) 1st shell, (b) 2nd shell, (c) 3rd shell, (d) 4th shell, and (e) 5th shell. 
The number in (c) indicates the $i$th Yb site and the number in (b), (d), and (e) indicates the $\xi$th Al site.   
}
\label{fig:Tsai_cluster}
\end{figure}

The quasicrystal consists of the Tsai-type cluster which has concentric shell structures of 1st, 2nd, 3rd, 4th and 5th shells, as shown in Fig.~\ref{fig:Tsai_cluster}. In the 3rd shell, which is icosahedron, 12 Yb atoms form the vertexes. In the 1st, 2nd, and 4th shells, Al/Au mixed sites exist, whose existence ratios are 7.8$\%$/8.9$\%$, 62$\%$/38$\%$, and 59$\%$/41$\%$, respectively~\cite{Ishimasa}. 
There also exists the 1/1 AC Yb$_{14}$Al$_{35}$Au$_{51}$, which has periodic arrangement of 
the Tsai-type cluster forming the body-centered-cubic (bcc) structure. 
In the $F_{n-1}/F_{n-2}$ AC, where $F_n$ is Fibonacci number satisfying the relation $F_n=F_{n-1}+F_{n-2}$ for $n\ge 3$ with $F_1=1$ and $F_2=1$, 
as $n$ increases, the size of the unit cell increases and the $n\to\infty$ limit corresponds to the QC~\cite{Goldman1993}. 

%%%%%%%%%%%%%%%  Fig.2 %%%%%%%%%%%%%%%%%%%%%%%%%%%%
%%%%%%%%%%%%%%%%%%%%%%%%%%%%%%%%%%%%%%%%%%%%%%%%%%%

\subsection{Lattice constant dependence of Yb valence in quasicrystal Yb$_{15}$(Al, Ga)$_{34}$(Au, Cu)$_{51}$}
\label{sec:Lattice_QC}

\subsubsection{Model and method}
\label{sec:MM}

To get insight into the mechanism of the measured lattice-constant dependence of the Yb valence in the QC Yb$_{15}$(Al$_{1-x}
$Ga$_{x}$)$_{34}$(Au$_{1-y}$Cu$_{y}$)$_{51}$, we perform the theoretical analysis by constructing the minimal model in the AC. 
The minimal model in the hole picture consists of the 4f and 5d orbitals at Yb and the 3p orbital at Al: 
\begin{eqnarray}
H&=&
\sum_{\langle j{\xi}, j'{\nu}\rangle\sigma}
\left(t_{j{\xi}, j'{\nu}}^{\rm pp}
c_{j\xi\sigma}^{\dagger}c_{j'\nu\sigma}+{\rm h.c.}
\right)
+\sum_{\langle ji, j'i'\rangle\sigma}
\left(t_{ji, j'i'}^{\rm dd}
d_{ji\sigma}^{\dagger}d_{j'i'\sigma}+{\rm h.c.}
\right)
+\sum_{\langle ji, j'\xi\rangle\sigma}
\left(V_{ji, j'\xi}^{\rm dp}
d_{ji\sigma}^{\dagger}c_{j'\xi\sigma}+{\rm h.c.}
\right) 
\nonumber
\\
&+&
\sum_{\langle ji, j'\xi\rangle\sigma}
\left(V_{ji, j'\xi}^{\rm fp}
f_{ji\sigma}^{\dagger}c_{j'\xi\sigma}+{\rm h.c.}
\right)
+\sum_{j=1}^{N_{\rm L}}
\left[
\varepsilon_{\rm f}\sum_{i=1\sigma}^{24}n_{ji\sigma}^{\rm f}
+U\sum_{i=1}^{24}n_{ji\uparrow}^{\rm f}n_{ji\downarrow}^{\rm f}
\right]
+U_{\rm fd}\sum_{j=1}^{N_{\rm L}}\sum_{i=1}^{24}n_{ji}^{\rm f}n_{ji}^{\rm d},  
\label{eq:EPAM_AC}
\end{eqnarray}
where the 1st and 2nd terms are transfers of 3p holes between the Al sites and 5d holes between the Yb sites, respectively. 
The 3rd (4th) term is the hybridization between Yb 5d (4f) hole and Al 3p hole. 
In the 5th term, $\varepsilon_{\rm f}$ is the energy level of 4f hole and $U$ is the onsite Coulomb repulsion between the 4f holes at Yb. 
The last term is the onsite Coulomb repulsion between the 4f and 5d holes at Yb, where $n^{\rm a}_{ji}$ for ${\rm a}={\rm f}, {\rm d}$ is defined as $n^{\rm a}_{ji}\equiv n^{\rm a}_{ji\uparrow}+n^{\rm a}_{ji\downarrow}$ with $\uparrow$ and $\downarrow$ being the Kramers pairs in the CEF ground state.
Since the bcc unit cell contains two icosahedrons, there exist 24 Yb sites in the unit cell, which result in the index $i=1$-$24$ for 4f and 5d operators in Eq.~(\ref{eq:EPAM_AC}) [see Fig.~\ref{fig:Tsai_cluster}(c)]. 
%%%%%
%\textcolor{red}
%{
The index $\xi=1$-$48$ denotes the Al site as shown in Figs.~\ref{fig:Tsai_cluster}(b), \ref{fig:Tsai_cluster}(d), and \ref{fig:Tsai_cluster}(e). 
%}
%%%%%
The index $j$ specifies the unit cell and $N_{\rm L}$ is the total number of unit cells. 
  
As a first step of analysis, we consider the case that Al/Au mixed sites are occupied by Al and neglect orbital degeneracy. 
By substituting the lattice distances into the Harrison's formula~\cite{Harrison},  
the p-p and d-d transfers between the nearest-neighbor (N.N.) Al and Yb sites and the d-p hybridization between the N.N, Yb and Al sites are estimated as $t^{\rm pp}\approx 4.0$~eV, $t^{\rm dd}\approx 0.03$~eV, and $V^{\rm dp}\approx 0.42$~eV, respectively (see Ref.~\cite{WM2018} for detail). 
The other transfers and hybridizations between the shells are set so as to follow  
%$t_{ji,j'i'}^{\rm pp(dd)}\propto 1/r^{\ell+\ell'+1}$, 
%%%%%
%\textcolor{red}
%{
$ t^{\rm pp}_{j\xi,j'\xi'} (t^{\rm dd}_{ji,j'i'})\propto 1/r^{\ell+\ell'+1}$
%}
%%%%%
where 
$r$ is the distance between the 3p (5d) orbitals with azimuthal quantum numbers, $\ell=1~(2)$ and $\ell'=1~(2)$, respectively, and 
$V_{ji,j'\xi}^{\rm f(d)p}\propto 1/r^{\ell+\ell'+1}$, where $r$ is the distance between the 4f (5d) orbital and the 3p orbital with azimuthal quantum numbers, $\ell=3~(2)$ and $\ell'=1$, respectively,  
following the linear muffin-tin orbital (LMTO) argument~\cite{Andersen1}. 

To analyze the heavy-fermion state, we apply the slave-boson mean-field theory for $U=\infty$ to the Hamiltonian [Eq.~(\ref{eq:EPAM_AC})]~\cite{Read,OM2000}. As for the $U_{\rm fd}$ term in Eq.~(\ref{eq:EPAM_AC}), the mean-field approximation is applied~\cite{WM2018}. 
We performed the calculation to obtain the ground state of Eq.~(\ref{eq:EPAM_AC}) by varying $\varepsilon_{\rm f}$, $U_{\rm fd}$, and hybridization $V^{\rm fp}$ between the N.N. Yb on the 3rd shell and Al on the 2nd shell as input parameters at half filling. 
As for the 5d level $\varepsilon_{\rm d}$, we set $\varepsilon_{\rm d}=0$ in Eq.~(\ref{eq:EPAM_AC}) for simplicity since main result below is 
%expected not to change essentially by 
robust against the moderate change of 
the values of $\varepsilon_{\rm d}$ in the case that the 5d band has a certain filling. 
Below we take $t^{\rm pp}$ as the unit of energy and show the results for $V^{\rm fp}=0.1$ as a typical value in the $N_{\rm L}=8^3$ lattice system. 

\subsubsection{Results}

%\subsubsection{Ground-state phase diagram}

\begin{figure}[h]
\includegraphics[width=14cm]{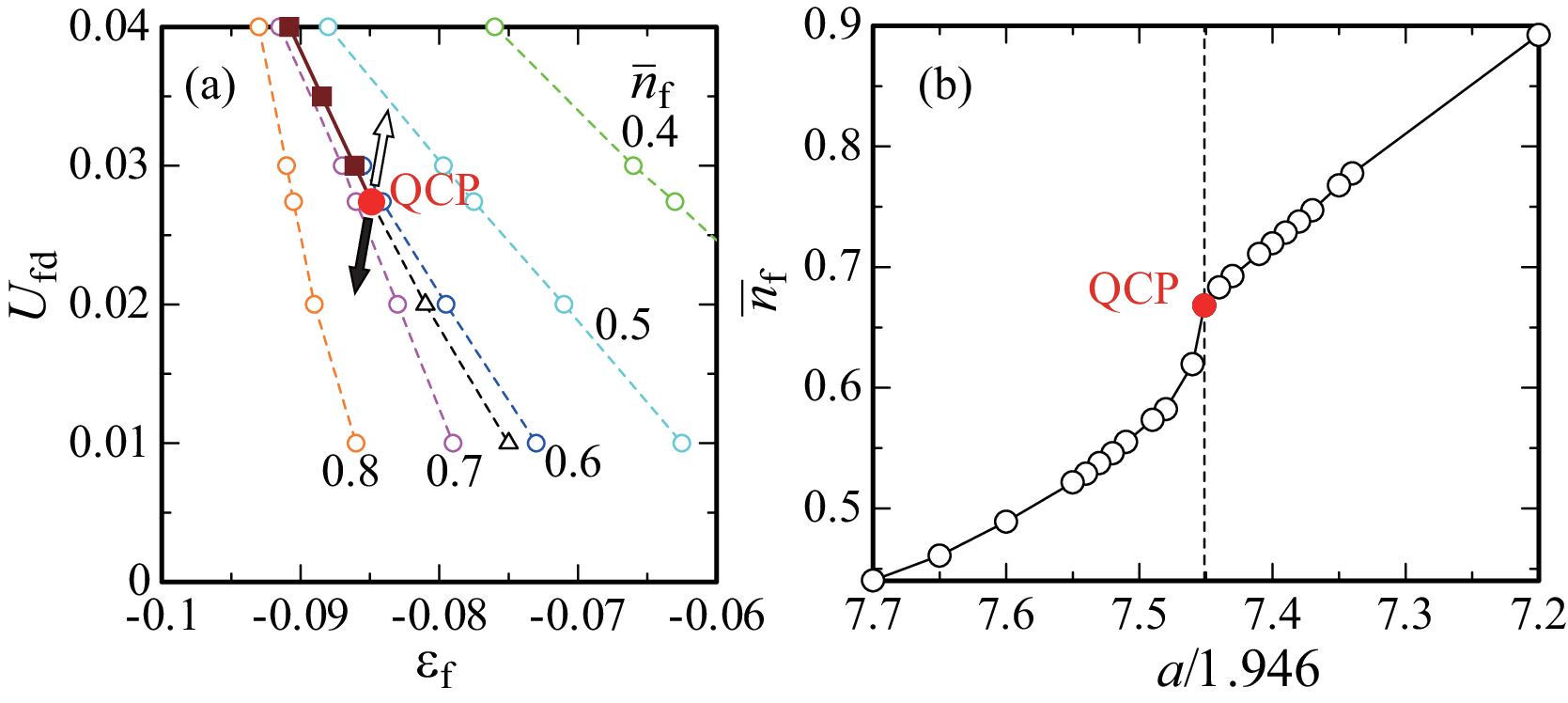}
\caption{(Color online) (a) Ground-state phase diagram of the Hamiltonian [Eq.~(\ref{eq:EPAM_AC})] in the $\varepsilon_{\rm f}$-$U_{\rm fd}$ plane for $V^{\rm fp}=0.1$ at half filling. 
First-order transition line (solid line with filled squares) terminates at the QCP of valence transition (filled circle) and sharp valence-crossover line extends from the QCP (dashed line with open triangles). 
Contour lines for $\bar{n}_{\rm f}=0.4$, $0.5$, $0.6$, $0.7$, and $0.8$ are also shown (dashed lines with open circles). 
Open (filled) arrow indicates direction of parameter change as lattice expands (shrinks) (see text). 
(b) Lattice constant dependence of 4f-hole number per Yb staring from the valence QCP (filled circle) in the AC with lattice constant $a=14.5$~$\AA$ at ambient pressure denoted by the vertical dashed line. 
}
\label{fig:PD_nf_a}
\end{figure}

We calculated the $\varepsilon_{\rm f}$ dependence of the 4f-hole number per Yb defined as 
$\bar{n}_{\rm f}\equiv\sum_{j=1}^{N_{\rm L}}\sum_{i=1}^{24}\langle n_{ji}^{\rm f}\rangle/(24N_{\rm L})$ for various $U_{\rm fd}$. 
Figure~\ref{fig:PD_nf_a}(a) shows the ground-state phase diagram in the $\varepsilon_{\rm f}$-$U_{\rm fd}$ plane. 
The first-order valence transition indicating a jump in $\bar{n}_{\rm f}$ (solid line with filled squares) terminates at the QCP (filled circle) and 
the sharp valence-crossover line extends from QCP (dashed line with open triangles). 
At the QCP realized for $(\varepsilon_{\rm f}^{\rm QCP}, U_{\rm fd}^{\rm QCP})=(-0.0849, 0.0274)$, critical valence fluctuations diverge $\chi_{\rm v}\equiv -\partial\bar{n}_{\rm f}/\partial\varepsilon_{\rm f}=\infty$.  
The value of the 4f-hole number per Yb is $\bar{n}_{\rm f}=0.67$. 
This corresponds to Yb$^{+2.67}$, which explains the measured intermediate valence of Yb~\cite{Watanuki,Imura}. 
We also plot contour lines for $\bar{n}_{\rm f}=0.4$, $0.5$, $0.6$, $0.7$, and $0.8$ (dashed lines with open circles). 
As $\varepsilon_{\rm f}$ decreases $\bar{n}_{\rm f}$ increases and $\bar{n}_{\rm f}=1$ is realized for deep $\varepsilon_{\rm f}$. Hence, the left (right) side of the first-order valence-transition line and valence-crossover line is called Kondo (mixed-valence) regime. 
Here we note that f-p hybridization is finite $\langle f_{ji\sigma}^{\dagger}c_{j'\xi\sigma}\rangle\ne 0$ everywhere in the phase diagram. 
Hence, quasiparticles of heavy fermions are formed everywhere while its mass and $\bar{n}_{\rm f}$ i.e., Yb valence depend on the location in the phase diagram.

%%%%%%%%%%%%%%%  Fig.3 %%%%%%%%%%%%%%%%%%%%%%%%%%%%
%%%%%%%%%%%%%%%%%%%%%%%%%%%%%%%%%%%%%%%%%%%%%%%%%%%

%\subsubsection{Lattice constant dependence of Yb valence}

Next, let us proceed to the analysis of the lattice constant dependence of Yb valence. 
In order to perform such a calculation, we need to know the lattice-constant dependence of the 4f level $\varepsilon_{\rm f}$. Generally, the origin of the CEF is ascribed to the electrostatic interaction and hybridization. Which effect is predominant is material dependent. As for TbCd$_6$, it was reported that in the point charge model the CEF parameter $B_{20}$ is essentially enough to explain the specific-heat~\cite{Jazbec2016} and neutron-scattering measurements~\cite{Das2017}. In this case, 4f level has $r^{-3}$ dependence. 

On the other hand, in the hybridization picture, 4f level can be evaluated as the second-order perturbation with respect to hybridization as a leading term $\varepsilon_{\rm f}\propto|V^{\rm fc}_{ji,j'\xi}|^2/\Delta$, which gives $r^{-10}$ dependence since the f-p hybridization has the 
$V^{\rm fc}_{ji,j'\xi}\propto r^{-5}$ dependence as mentioned above. 
Here, $\Delta$~$(>0)$ is the first-excited energy from the CEF ground state. 
In the QC Yb$_{15}$Al$_{34}$Au$_{51}$, to reproduce the measured increase in Yb valence under pressure~\cite{Watanuki,Imura}, it turned out that effect of hybridization is indispensable. So, here we employ $r^{-10}$ dependence in the 4f level $\varepsilon_{\rm f}$.

To clarify how the Yb valence changes in the vicinity of the valence QCP as a function of the lattice constant, we have investigated how $\bar{n}_{\rm f}$ changes starting from the QCP in the AC. 
First, we input parameters for the valence QCP to Eq.~(\ref{eq:EPAM_AC}) in the AC with the lattice constant at ambient pressure, $a=14.5$ ð. Then, by changing the lattice constant $a$, we calculate the 4f-hole number per Yb site $\bar{n}_{\rm f}$ without inputting any adjustable parameters. Namely, the lattice-constant dependence of transfers, hybridizations, and 4f level in Eq.~(\ref{eq:EPAM_AC}) are automatically taken into account and the white circles in Fig.~\ref{fig:PD_nf_a}(b) are obtained.  
Here we plot $x$-axis as the 6-dimensional lattice constant, which is transformed from 3D lattice constant as $a_{6{\rm D}}\equiv a/1.946$, corresponding to the ``lattice constant" of the QC~\cite{Imura}. 
The result shows sudden decrease from the valence QCP. This captures the experimental feature that the quantum-critical QC Yb$_{15}$Al$_{34}$Au$_{51}$ is located at the point where Yb valence starts to change sharply. The local structure around the Yb site is common in both the AC and QC and valence transition is caused by on-site 4f-5d Coulomb repulsion at Yb site, which is local. So, this result suggests that the QC Yb$_{15}$Al$_{34}$Au$_{51}$ is located at the valence QCP. 

%If our theory is correct, it should be able to predict future experiment. That is pressure dependence of Yb valence in the AC. 
According to our theory, the pressure dependence of Yb valence in the AC is predicted, which can be verified experimentally. 
At ambient pressure, the AC Yb$_{14}$Al$_{35}$Au$_{51}$ shows the Fermi-liquid behavior. By applying pressure, it shows the same quantum criticality as that in the quasicrystal at $P\approx 2$~GPa~\cite{Matsukawa2014}. So, this sharp change in Yb valence is expected to appear at 2 GPa. This prediction has actually been observed recently by Imura et al~\cite{Imura}. 

One remaining question is why sudden drop in Yb valence starts to appear from QCP in Fig.~\ref{fig:PD_nf_a}(b). To clarify this point, let us focus on the ground-state phase diagram in Fig.~\ref{fig:PD_nf_a}(a).

As the lattice shrinks, 4f hole level $\varepsilon_{\rm f}$ decreases. At the same time, 4f-3p hybridization increases because 4f orbital and 3p orbital approach. Then, location of the QCP shifts to left-upper direction in the $\varepsilon_{\rm f}$-$U_{\rm fd}$ plane, which is accompanied by the shift of the contour lines of 4f hole numbers. The relative position from the QCP can be redrawn from its original position, whose direction is illustrated schematically as an filled arrow in Fig.~\ref{fig:PD_nf_a}(b). 

On the other hand, as lattice expands, 4f hole level $\varepsilon_{\rm f}$ increases. At the same time, 4f-3p hybridization decreases. Then, location of the QCP shits to right-lower direction in the $\varepsilon_{\rm f}$-$U_{\rm fd}$ plane, which is accompanied by the shift of the contour lines of 4f hole numbers. The relative position from the QCP can be redrawn from its original position, whose direction is illustrated schematically as an open arrow in Fig.~\ref{fig:PD_nf_a}(b). 

Since the former direction indicated by the open arrow goes across much dense contour area than the latter direction indicated by the filled arrow in the opposite direction, the former gives larger $\bar{n}_{\rm f}$ change than the latter. 
This clarifies why $\bar{n}_{\rm f}$ decreases from the QCP sharply as $a$ increases from $a=14.5$~\AA~ denoted by the vertical dashed line in Fig.~\ref{fig:PD_nf_a}(b).  
This indicates that hybridization dependence of the location of the QCP causes asymmetric lattice constant dependence of the Yb valence, giving rise to the sudden decrease from the QCP. 

\subsection{$T/B$ scaling and $\chi(T)\sim T^{-0.5}$ in quasicrystal Yb$_{15}$Al$_{34}$Au$_{51}$}
\label{sec:AC_TB}

%%%%%
%\textcolor{red}
%{
The quasicrystal Yb$_{15}$Al$_{34}$Au$_{51}$ exhibits essentially the same $T/B$ scaling as well as $\chi(T)\sim T^{-0.5}$ behavior for the zero-field limit as observed in the periodic crystal $\beta$-YbAlB$_4$. These behaviors are explained on the basis of the Hamiltonian-based mode-coupling theory for the AC, which will be explained in Sect.~\ref{sec:TB_AC}. The key of the emergence of the $T/B$ scaling is shown to be the small characteristic temperature of the CVF, which will be discussed in Sect.~\ref{sec:small_T0_AC}.
%}
%%%%%

%%%%%
%\textcolor{red}
%{
\subsubsection{Analysis by the Hamiltonian-based mode-coupling theory}
\label{sec:TB_AC}
%}
%%%%%

\begin{figure}[tb]
\includegraphics[width=14cm]{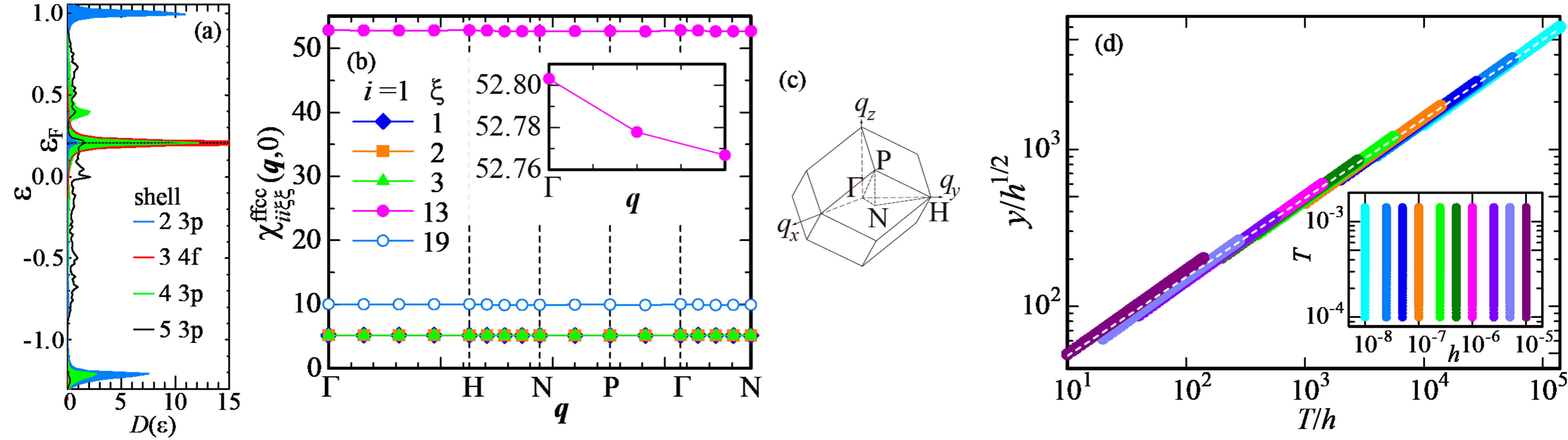}
\caption{(Color online) (a) Quasiparticle density of states of 4f electron (3rd shell) and 3p electron (2nd, 4th, and 5th shells) in the hole picture. Fermi level is located at $\varepsilon_{\rm F}$. 
(b) $\chi^{\rm ffpp}_{ii\xi\xi,\sigma}({\bf q}, 0)$ for $i=1$ and $\xi=1$ (filled diamond), 
2 (filled square), 3 (filled triangle), 13 (filled circle), and 19 (open circle) for $T=0.0001$. 
%Inset shows Brillouin zone of bcc lattice. 
%%%%%
%\textcolor{red}
%{
Inset shows enlargement of $\chi^{\rm ffcc}_{1,1,13,13}({\bf q},0)$ along the $\Gamma$-H line.
(c) Brillouin zone of bcc lattice.
(d) 
%}
%%%%%
Scaling plot of $y/h^{1/2}$ vs. $T/h$. Here, $y(T,h)$ is the solution of the valence SCR equation where parameters except for $T$ and $h$ are set as the values for the valence QCP at $h=0$. 
The dashed line represents the fitting function $c(T/h)^{\zeta}$ with $\zeta=0.503$ (see text).
Inset shows the data in the $T$-$h$ plane where the scaling applies~\cite{WM2016}.  
}
\label{fig:AC_TB}
\end{figure}

The $T/B$ scaling and $\chi(T)\sim T^{-0.5}$ observed in the QC Yb$_{15}$Al$_{34}$Au$_{51}$ are also explained from theoretical analysis on the basis of the periodic Anderson model in the AC Yb$_{14}$Al$_{35}$Au$_{51}$~\cite{WM2016}. 
By applying the slave-boson mean-field theory for $U=\infty$ as described in Sect.~\ref{sec:MM},  
we obtained the ground state. 
We calculated the band structures for various parameters and the result for $\varepsilon_{\rm f}=-0.4$ corresponding to the pressurized AC is shown here (see Ref.~\cite{WM2016} for detail). 
Figure~\ref{fig:AC_TB}(a) shows the density of states of the quasiparticles. 
%%%%%
%\textcolor{red}
%{
In the plot we set the imaginary part of the self-energy quasiparticle as $\delta=0.01$  
%%%%%
%\textcolor{blue}
%{
in order to show the overall feature of the density of states with smooth $\varepsilon$ dependence not affected by finite-size effects in the $N_{\rm L}=8^3$ lattices. 
%}
%%%%%

We note here that the renormalization factor for this state is $z=1.9\times 10^{-3}$, 
%which is three decades of smaller than the conduction electron band width. 
%%%%%
%\textcolor{blue}
%{
which is obtained directly from the solution $\bar{b}_i$ of the slave-boson mean-field equations as $z\equiv\bar{b}_{i}^2$ (see Ref.~\cite{WM2016} for detail). Since this small $z$ implies that the Kondo temperature, which is three decades smaller than the conduction-electron band width, is realized in the present lattice system, this state is in the heavy-fermion regime. 
%}
%%%%%
%Hence, this state is in the heavy-fermion regime.
%}
%%%%% 
The dominant contribution to the density of states at the Fermi level comes from 4f hole at Yb site and the next leading contribution comes from 3p hole at Al site on the 4th shell. This is due to the strongest f-p hybridization reflecting the shortest Yb-Al distance [see Figs.~\ref{fig:Tsai_cluster}(c) and \ref{fig:Tsai_cluster}(d)]. 

By employing the obtained quasiparticle state as a solution of the slave-boson mean-field equations, we calculate the dynamical susceptibility for the charge transfer between 4f and conduction holes 
\begin{eqnarray}
\chi^{\rm ffpp}_{ii\xi\xi,\sigma}({\bf q},i\omega_{m})=
-\frac{T}{N_{\rm L}}\sum_{n{\bf k}}G_{ii,\sigma}^{\rm ff}({\bf k}+{\bf q},i\varepsilon_{n}+i\omega_{m})
G_{\xi\xi,\sigma}^{\rm pp}({\bf k},i\varepsilon_{n}), 
\label{eq:chi_ffcc}
\end{eqnarray}
where $G^{\rm ff}_{ii,\sigma}$ and $G^{\rm pp}_{\xi\xi,\sigma}$ are f and p components of the quasiparticle Green function $\hat{G}_{\sigma}({\bf k},i\varepsilon_{n})\equiv(i\varepsilon_{n}-\tilde{H}_{{\bf k}\sigma}+\mu)^{-1}$, 
where $\tilde{H}_{{\bf k}\sigma}$ is the mean-field Hamiltonian  
with 
$\varepsilon_{n}=(2n+1)\pi{T}$ and the chemical potential $\mu$, and $\omega_{m}=2m\pi{T}$. 

%%%%%%%%%%%%%%%  Fig.4 %%%%%%%%%%%%%%%%%%%%%%%%%%%%
%%%%%%%%%%%%%%%%%%%%%%%%%%%%%%%%%%%%%%%%%%%%%%%%%%%

Figure~\ref{fig:AC_TB}(b) shows the momentum dependence of $\chi^{\rm ffpp}_{ii\xi\xi,\sigma}({\bf q}, 0)$ calculated for $T=0.0001$ with the number of Matsubara frequency $2^{15}$ in $N_{\rm L}=8^3$. 
Here,  
%%%%%
%\textcolor{red}
%{
Fig.~\ref{fig:AC_TB}(c) 
%}
%%%%%
shows the Brillouin zone of the bcc lattice of the AC. 
The index $i$ and $\xi$ specify the Yb and Al sites respectively shown in Fig.~\ref{fig:Tsai_cluster}. 
We find that the charge transfer mode between 4f hole at Yb on the 3rd shell and 3p hole at Al on the 4th shell is enhanced [e.g. the mode for $i=1$ and $\xi=13$ is the largest, see Figs.~\ref{fig:Tsai_cluster}(c) and \ref{fig:Tsai_cluster}(d)]. 
A remarkable result is almost flat momentum dependence appears, which is ascribed to strong local correlation effect for 4f holes at Yb. 
To clarify how the local charge-transfer fluctuation affects the quantum criticality, let us focus on the largest charge-transfer mode [between N.N. Yb and Al sites, e.g. the mode for $i=1$ and $\xi=13$ in Fig.~\ref{fig:AC_TB}(b)] because it is overwhelmingly dominant.
%%%%%
%\textcolor{red}
%{
We note that $\chi_{1,1,13,13}^{\rm ffcc}({\bf q},0)$ has a maximum at the $\Gamma$ point ${\bf q}=(0,0,0)$ [see inset in Fig.~\ref{fig:AC_TB}(b), which shows the enlargement along the $\Gamma$-H line]. 
%}
%%%%%

Then, we apply the recently-developed mode-coupling theory of critical valence fluctuations under magnetic field~\cite{WM2014} to the present system. Namely, starting from the periodic Anderson model with the inter-orbital Coulomb repulsion and Zeeman term 
\begin{eqnarray}
H_{\rm Zeeman}=
-h\sum_{j=1}^{N_{\rm L}}
\left[\sum_{i=1}^{24}S_{ji}^{{\rm f}z}
+\sum_{\xi=1}^{48}S_{j\xi}^{{\rm c}z}
\right],  
\label{eq:H_Zeeman}
\end{eqnarray}
where $h$ is magnetic field, and $S_{ji}^{{\rm f}z}\equiv\frac{1}{2}(n_{ji\uparrow}^{\rm f}-n_{ji\downarrow}^{\rm f})$ 
and $S_{j\xi}^{{\rm c}z}\equiv\frac{1}{2}(n_{j\xi\uparrow}^{\rm c}-n_{j\xi\downarrow}^{\rm c})$, 
we have derived the valence SCR equation (see Ref.~\cite{WM2016} for detail). By solving the valence SCR equation at the valence QCP corresponding to the pressure-tuned AC, we found that $T/h$ scaling appears over four decades, as shown in Fig.~\ref{fig:AC_TB}(d). 
The dashed line indicates the least-square fit of the data by the scaling function 
\begin{eqnarray}
y=h^{1/2}\phi\left(\frac{T}{h}\right).
\label{eq:AC_scaling}
\end{eqnarray}
Here, $\phi(x)$ is given by $\phi(x)=cx^{\zeta}$ 
in the large-$T/h$ region for $T/h>10^2$, which gives the exponent $\zeta=0.503$. Then, by the relation $y/h^{0.5}=c(T/h)^{0.5}$, the solution of the valence SCR equation $y$ turns out to have $y\propto T^{0.5}$ dependence. So, magnetic susceptibility as well as valence susceptibility shows $T^{-0.5}$ criticality, $\chi_{\rm v}\propto\chi\propto T^{-0.5}$.

%%%%%
%\textcolor{red}
%{
For $T/h\lsim 10^2$, the data starts to deviate from the dashed line in Fig.~\ref{fig:AC_TB}(d). This indicates the crossover to the Fermi-liquid regime for $T/h\ll 1$ starts to appear, where the magnetic susceptibility $\chi(T,h)$ at low $T$ is suppressed under the magnetic field. This tendency is also consistent with the experiments~\cite{DS_pc,Matsukawa2014}. 
%}
%%%%%

%This 
%%%%%
%\textcolor{red}
%{
The result of Eq.~(\ref{eq:AC_scaling}) 
%}
%%%%%
indicates that the magnetic susceptibility in the pressurized AC exhibits 
\begin{eqnarray}
\chi\propto T^{-0.5}
\label{eq:chi_AC}
\end{eqnarray}
for zero-field limit and also the $T/B$ scaling behavior as observed in the QC. 
The locality of critical valence fluctuations shown in Fig.~\ref{fig:AC_TB}(b) is the origin of the emergence of the quantum valence criticality. 
Since the local structure around Yb is common in both QC and AC (see Fig.~\ref{fig:Tsai_cluster}) and the infinite limit of the unit-cell size of the AC corresponds to the QC, these results of Eqs.~(\ref{eq:AC_scaling}) and (\ref{eq:chi_AC}) are considered to capture the essence of the quantum criticality in the QC Yb$_{15}$Al$_{34}$Au$_{51}$. 
Actually, the $T/B$ scaling as well as $\chi(T)\sim T^{-0.5}$ is observed in the AC at $P=1.96$~GPa, which verifies our theory~\cite{Matsukawa2014}.  

\subsubsection{Small characteristic temperature of critical valence fluctuations}
\label{sec:small_T0_AC}

%In the calculation in Sect.~\ref{sec:AC_TB}, we evaluate the characteristic temperature of critical valence fluctuations as $T_0=1.4\times 10^{-4}$ in the unit of $t^{\rm pp}$, which is located around the lowest temperature [see the inset in Fig.~\ref{fig:AC_TB}(c)]. 
%%%%%
%\textcolor{red}
%{
In Fig.4(a), $\chi_{ii\xi\xi}^{\rm ffcc}({\bf q},0)$ for $i=1$ and $\xi=13$ has a maximum at the $\Gamma$ point ${\bf q}=(0,0,0)$ [see inset in Fig.4(b)]. We evaluated the dispersion of the most dominant charge-transfer mode, i.e., the ${\bf q}^2$ coefficient in 
\begin{eqnarray}
\chi_{ii\xi\xi}^{\rm ffcc}({\bf q}_{\nu},0)=\chi_{ii\xi\xi}^{\rm ffcc}({\bf 0},0)-A_{\nu}{\bf q}_{\nu}^2,
\end{eqnarray}
where ${\bf q}_{\nu}=\frac{2\pi}{a}(\frac{2}{N_1},0,0)$, $\frac{2\pi}{a}(0,\frac{2}{N_2},0)$, $\frac{2\pi}{a}(0,0,\frac{2}{N_3})$, $\frac{2\pi}{a}(\frac{1}{N_1},\frac{1}{N_2},0)$, $\frac{2\pi}{a}(0,\frac{1}{N_2},\frac{1}{N_3})$, $\frac{2\pi}{a}(\frac{1}{N_1},0,\frac{1}{N_3})$ 
with $a=14.5$~$\AA$ \ being a lattice constant in the $N_{\rm L}=N_1N_2N_3$ system.
%}
%%%%%

%%%%%
%\textcolor{red}
%{
We also evaluated the frequency dependence
%%%%%
\begin{eqnarray}
\chi^{\rm ffcc}_{ii\xi\xi}({\bf q}_{\nu},i\omega_{l})-\chi_{ii\xi\xi}^{\rm ffcc}({\bf q}_{\nu},0)\approx C_{\nu}\frac{\omega_{l}}{q_{\nu}}. 
\end{eqnarray}
%%%%%
By averaging the obtained coefficients over the momentum direction around the $\Gamma$ point $A_{\nu}$ and $C_{\nu}$ as $A_{\rm av}=\sum_{\nu=1}^{6}A_{\nu}/6$ and $C_{\rm av}=\sum_{\nu=1}^{6}C_{\nu}/6$, respectively, we obtained the characteristic temperature of CVF defined by 
\begin{eqnarray}
T_0\equiv\frac{A_{\rm av}q_{\rm B}^3}{2\pi C_{\rm av}},
\end{eqnarray}
where $q_{\rm B}$ is the momentum at the Brillouin zone, which is set to be $q_{\rm B}=(2\pi)/a$. The resultant $T_0$ is obtained as $T_0=1.4\times 10^{-4}$ in the unit of $t^{\rm pp}$. Hence, the localness of the CVF is reflected in the smallness of $T_0$, which is located around the lowest temperature [see the inset in Fig.~\ref{fig:AC_TB}(d)]. We note that $T_0$ is smaller than the Kondo temperature, which is in the order of $10^{-3}t^{\rm pp}$. 
%}
%%%%%
This extremely small $T_0$ is a consequence of locality of critical valence fluctuations characterized by the almost dispersionless mode shown in Fig.~\ref{fig:AC_TB}(b). 
Emergence of $T/B$ scaling is ascribed to the small characteristic temperature $T_0$, 
as discussed in Sect.~\ref{sec:small_T0}. 

In the QC, $T_0$ is expected to be smaller than the measured lowest temperature $T=30$~mK~\cite{WM2016}. In that case, the measured resistivity $\rho(T)\sim T$ down to the lowest temerature (see Table~\ref{tb:QCP}) is naturally explained from the theory of critical valence fluctuations.  
To examine this point, the M{\"o}ssbauer measurement in the QC Yb$_{15}$Al$_{34}$Au$_{51}$ to detect time scale of valence fluctuations $\tau$ providing a clue to $T_0$ as done in $\beta$-YbAlB$_4$ is highly desirable as future experiments. 

\subsection{Robust criticality in QC Yb$_{15}$Al$_{34}$Au$_{51}$ under pressure}
\label{sec:robustP}

The quantum criticality in the QC Yb$_{15}$Al$_{34}$Au$_{51}$ emerges at ambient pressure and surprisingly it persists even under pressure at least up to $P=1.6$~GPa~\cite{Deguchi}. 
To clarify the origin and mechanism, 
theoretical analysis was performed on the basis of the extended Anderson model on the Tsai-type cluster shown in Fig.~\ref{fig:Tsai_cluster}~\cite{WM2013}. 
%Because of the Al/Au mixed sites in the Tsai-type cluster, Yb sites in the 3rd shell become inequivalent each other, since the f-p hybridization at each Yb site becomes different, which makes many QCPs of valence transition in the ground-state phase diagram. 
Because of the Al/Au mixed sites in the Tsai-type cluster, the f-p hybridization at each Yb site becomes different. Hence, Yb sites in the 3rd shell become inequivalent each other, which makes many QCPs of valence transition in the ground-state phase diagram. 
An important result is that valence QCPs appear as an aggregate of spots, whose critical regions are overlapped to be unified, giving rise to a wide quantum critical region. 
Since the QC corresponds to the infinite limit of the unit-cell size of the AC, valence QCPs are expected to appear as widespread and condensed spots like the Andromeda Galaxy in the bulk limit of the QC. 
When we apply pressure to Yb compounds, 4f-hole level decreases in general. So, in case that applying pressure follows the line located in the enhanced critical-valence-fluctuation region, robust quantum criticality appears under pressure. Emergence of a wide critical region gives a natural explanation for the reason why quantum criticality appears without tuning control parameters in the QC Yb$_{15}$Al$_{34}$Au$_{51}$. 

%----------------------------------------------------------------------------------------
\section{Summary and Perspective}
\label{sec:SP}

We have discussed that 
recent discoveries of direct evidence of the sharp Yb-valence changes with the quantum valence criticality 
in $\alpha$-YbAl$_{1-x}$Fe$_x$B$_4$~$(x=0.014)$ and the QC Yb$_{15}$(Al$_{1-x}
$Ga$_{x}$)$_{34}$(Au$_{1-y}$Cu$_{y}$)$_{51}$ $(x=y=0)$
as well as their theoretical explanations based on the low-energy Hamiltonians in each system  
give firm grounds that the valence QCPs are realized 
in the periodic crystal $\beta$-YbAlB$_4$ and the QC Yb$_{15}$Al$_{34}$Au$_{51}$. 
An important point is that the quantum valence criticality originates from locality of critical valence fluctuations with extremely small characteristic temperature $T_0$, which emerges irrespective of  geometry of lattices such as periodic or quasi-periodic crystals. 
The above studies have shown an existence of the new universality class of the unconventional quantum criticality 
in strongly correlated electron systems. 
There is a possibility that the QCP of valence transition with the quantum valence criticality is realized in other systems on periodic and aperiodic crystals. 

As one of such candidates, the critical point of the first-order valence transition has been identified to be $(T_{\rm  c}, P_{\rm c}, B_{\rm c})=(2~{\rm K}, 8.5~{\rm GPa}, 0.65~{\rm T})$ in the $T$-$P$-$B$ phase diagram of YbNi$_3$Ga$_9$~\cite{Matsubayashi}. 
On cooling at $(P, B)\approx (P_{\rm c}, B_{\rm c})$, enhancement of the uniform magnetic susceptibility $\chi(T)$ toward the critical point was also observed. 
It is interesting to identify the QCP at $T=0$~K by controlling pressure and magnetic field on cooling and then to observe temperature dependences of $\chi(T)$, $C(T)/T$, $\rho(T)$, and $(T_1T)^{-1}$, 
which examines the quantum valence criticality shown in Table~\ref{tb:QCP}. 
Such measurements in periodic and aperiodic systems are interesting future subjects.

\section*{Acknowledgments}
We acknowledge K. Kuga, Y.~Matsumoto, H. Kobayashi, K. Imura, K. Deguchi, S. Matsukawa, N.~K.~Sato, T.~Watanuki, and S. Nakatsuji for showing us experimental data with enlightening discussions. 
This work was supported by JSPS KAKENHI Grant Numbers JP18K03542, JP18H04326, and JP17K05555.

%--------------------------------------------------------------------------------------------

\end{document}